\documentclass[12pt]{iopart}
 \expandafter\let\csname equation*\endcsname\relax
 \expandafter\let\csname endequation*\endcsname\relax
\usepackage{amsmath,amsfonts,amssymb}
\usepackage{graphicx, caption}
\usepackage{setspace}
\usepackage{tocloft}
\usepackage{lineno}

\usepackage{dsfont}

\def\BE{\begin{equation}}
\def\EE{\end{equation}}
\def\BEA{\begin{eqnarray}}
\def\EEA{\end{eqnarray}}
\def\BY{\begin{eqnarray}}
\def\EY{\end{eqnarray}}
\def\BA{\begin{aligned}}
\def\EA{\end{aligned}}

\def\L{\label}

\def\({\left (}
\def\){\right )}
\def\[{\left [}
\def\]{\right]}
\def\<{\langle}
\def\>{\rangle}

\def\2!{\!\!}
\def\3!{\!\!\!}
\def\4!{\!\!\!\!}
\def\5!{\!\!\!\!\!}
\def\6!{\!\!\!\!\!\!}

\begin{document}

\title[]{Influence of the radial number of Laguerre-Gaussian vortex beams on their propagation in a turbulent medium}

\author{D. D. Reshetnikov$^{1*}$, A. S. Losev$^{1,2,3}$}

\address{$^1$Saint-Petersburg State University, Universitetskaya nab. 7/9, Saint-Petersburg, 199034, Russia}
\address{$^2$Saint Petersburg State Marine Technical University, Lotsmanskaya str. 3, Saint Petersburg, 190121, Russia}
\address{$^3$ITMO University, Kronverksky Prospekt 49, Saint Petersburg, 197101, Russia}

\ead{$^*$d.reshetnikov@spbu.ru}

\begin{abstract}
The paper considers the propagation of Laguerre-Gaussian beams in a turbulent gas medium. As demonstrated by numerical modelling and the decomposition of Laguerre-Gaussian modes according to the basis of orthogonal Zernike polynomials, disparities in the stability of Laguerre-Gaussian modes with differing radial and azimuthal numbers are evident. The demonstration is made of the minimization of the distortion of the transverse beam profile with an increase in the radial number.
\end{abstract}

%
\vspace{2pc}

\section{Introduction}
\label{sect:intro}  
The study of beams with spatial structure is of interest for a wide range of tasks, including ultra-high resolution microscopy \cite{Hell-1994}, the creation of optical traps \cite{McDonald-2002, Jesacher-2004}, as well as the encoding and transmission of information in classical \cite{Willner-2021, Trichili-2019, Lavery-2017, Doster-2016, Forbes-2016} and quantum \cite{Willner-2021, Trichili-2019, Mirhosseini-2019, Wang-2019} cases. The primary impediment to the extensive utilisation of optical information communication channels in free space pertains to the degradation of transmitted spatial modes in a turbulent atmosphere. The description of these phenomena has evolved significantly since the second half of the 20th century \cite{Tatarski-1961, Andrews-1992, Cheng-2009, Cox-2016, Lavery-2017, Lochab-2018, Cox-2020}. Among studies on the stability of vectorial structured light in a turbulent medium, there are those that suggest that the transverse profile may be stable \cite{Korotkova-2009, Cheng-2009, Lochab-2018, Klug-2023}, and those that suggest that it is not \cite{Ndagano-2017, Cox-2016}. 

Laguerre-Gaussian beams are of particular interest. Firstly, they are invariant with respect to propagation in a free, non-turbulent atmosphere and transformations in lenses \cite{Khorin-2018}. Secondly, they are one of the possible representations of beams with orbital angular momentum (OAM), which makes them convenient for encoding information in open optical communication channels \cite{Gibson-2004, Wang-2012, Willner-2015, Willner-2021} and free space quantum key distribution systems \cite{Golubeva-2024}.

In order to mathematically describe such communication channels, it is necessary to take into account errors associated with random perturbations of the medium in which the optical beam propagates \cite{Gladysz-2014, Lukin-2021, Lukin-2012}. For instance, there are well-known works devoted to the study of the behaviour of beams with OAM in a turbulent environment, as well as the influence of turbulence on their main characteristics \cite{Aksenov-2016, Aksenov-2012, Klug-2023}. Furthermore, as demonstrated in \cite{Aksenov-2012, Petrov-2024}, the utilisation of OAM beams has been shown to enhance the resistance of beams to fluctuations in the direction of propagation within turbulent media.

The wavefront analysis of such beams can be carried out by decomposing orthogonal Zernike polynomials \cite{Kirilenko-2016}. Zernike polynomials are frequently employed to analyse wavefront distortions induced by imperfections in optical components \cite{Wada-2005}. These functions are advantageous for the representation of wavefronts in numerical modelling of optical systems. In addition to providing a stable basis for errors associated with the accumulation of errors in calculations, they possess several salient features. Primarily, their decomposition coefficients are indicative of the presence of classical aberrations and can be easily measured experimentally with using Shack–Hartmann wavefront sensor. Moreover, their behaviour is well understood. It has been established that Zernike modes behave in turbulent media \cite{Khorin-2018}. Furthermore, it has been demonstrated that these polynomials, in addition to Laguerre-Gaussian (LG) modes, possess circular symmetry \cite{Wunsche-2005}.

In this paper, we demonstrate that certain LG modes exhibit minimal values of decomposition coefficients within the Zernike polynomial basis. This property renders them more resilient to turbulent atmospheric conditions. The results are then compared with numerical simulations of the passage of the corresponding LG beams through a turbulent atmosphere.

The paper is organized as follows. Section two discusses in detail the key features of LG modes and their distribution within a turbulent atmosphere. Section three examines the potential rationale behind the differential impact of a turbulent medium on different LG modes, based on their radial $p$ and azimuthal $l$ numbers. This section also introduces the decomposition of LG modes using Zernike polynomials. Section four presents an analysis based on this decomposition, as well as an evaluation of the stability of LG modes when propagating through a turbulent medium. Appendix provides a brief mathematical explanation of a technique for numerically simulating beam propagation in a turbulent environment.

\section{Methods: The modeling of the optical properties of turbulence}

The expression that describes the distribution of the LG beam amplitude in the transverse plane as a function of the longitudinal coordinate, $z$, at a fixed time, has the following form \cite{Solimenko-1989}:
\BE
LG_{p,\ell}(\rho,\varphi,z) =  \frac{\mathcal{N}_{p,\ell}}{w_z} \( \frac{\rho \sqrt{2}}{w_z} \)^{\2! |\ell|} \!
L^{|\ell|}_p \! \[ \frac{2 \rho^2}{w_z^2} \] e^{ - \frac{\rho^2}{w_z^2} (1 - i \frac{z}{z_R})}
e^{-i (2p + |\ell| + 1) \, \phi_g(z)} e^{i \, \ell \, \varphi}. \L{1}
\EE
Here $\rho,\varphi,z$ --- coordinates of the cylindrical coordinate system; $p\in\mathds{Z}\ge 0$ and $\ell\in\mathds{Z}$ --- radial and azimuthal numbers, respectively; $\mathcal{N}_{p,\ell} = \sqrt{ 2 \, p! / \( \pi \, (p + |\ell|)! \) }$ --- normalization constant; $z_R$ --- Rayleigh length \cite{Solimenko-1989}; $w_z^2 = w_0^2 \(1 + z^2 / z_R^2\)$ --- the square of the beam radius at a distance of $z$ from constrictions ($w_0 \equiv w_{z=0}$); $\phi_g(z) = \text{arctg}\(z/z_R\)$ --- Gouy phase; $L^{|\ell|}_p$ --- generalized Laguerre polynomial of order $p$ of degree $|\ell|$.

The numbers $\ell$ and $p$ are responsible for the shape of the transverse spatial profile of the beam, which is a set of $p+1$ concentric light rings and $p$ dark rings \cite{Solimenko-1989}; $\ell$ shows how many times the phase delay occurs on $2\pi$ in the transverse plane of the beam with one full rotation around the propagation axis $z$. The sign at $\ell$ corresponds to the direction of rotation. If the azimuthal number is different from zero, then there is a singularity in the center of the beam, that is, the intensity on the axis is zero.

In the context of atmospheric optical communication channels, LG beams are utilised for the transmission of information, characterised by distinct OAM values. One of the options for detecting the resulting state is to use a CCD matrix, which is capable of showing the result of the interference of the incoming LG beam with the reference Gaussian beam at a small angle between them. In this instance, the simulation was conducted with an angle of 1 mrad (see Fig.~\ref{Fig1}). A "fork" in the interference pattern will be observed, the size and shape of which are contingent on the magnitude of the azimuthal number, denoted by $\ell$.

\begin{figure}[h!]
\begin{minipage}[h]{0.24\linewidth}
\centering
\includegraphics[width=0.9\linewidth]{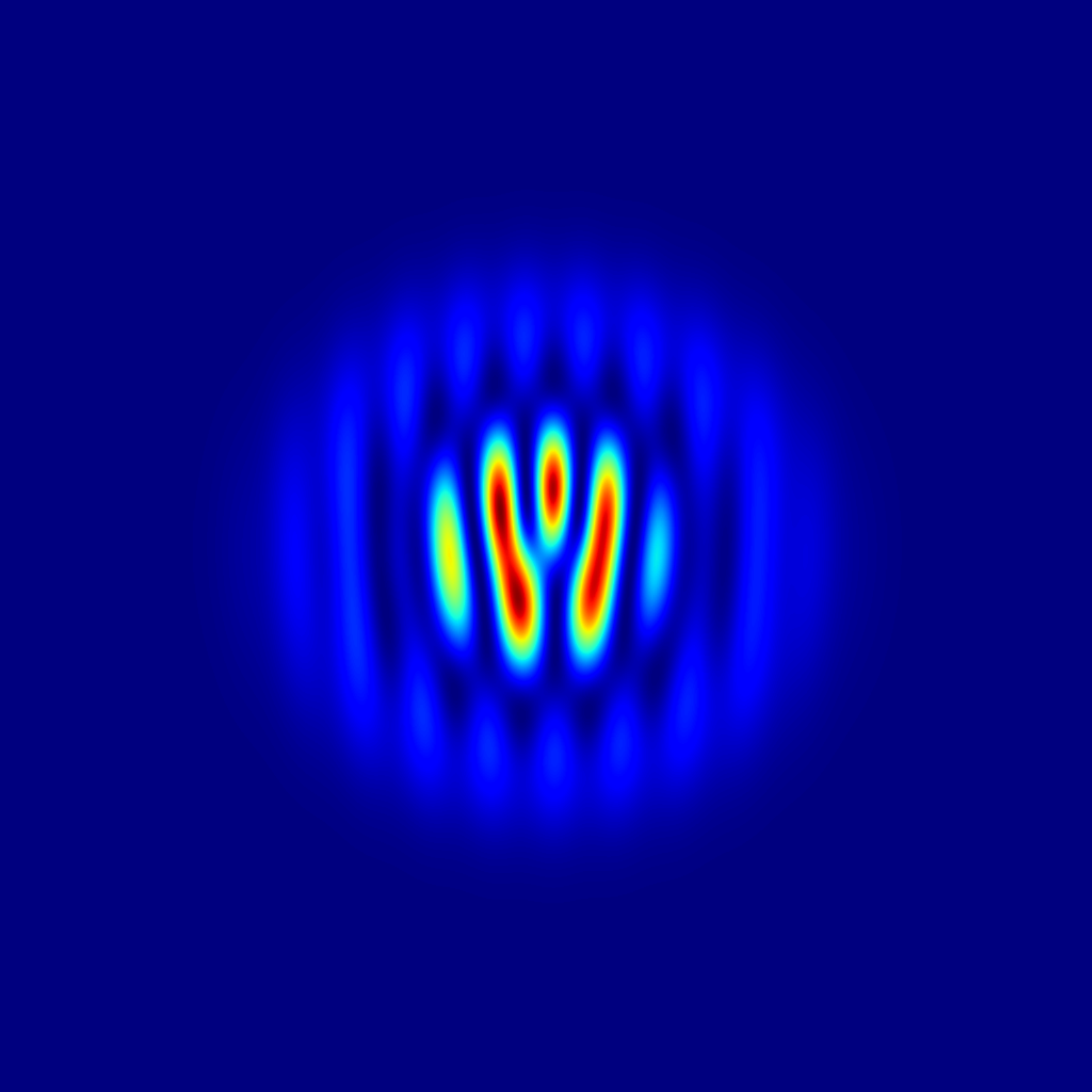} \\ $LG_{0,1}$
\end{minipage}
\hfill
\begin{minipage}[h]{0.24\linewidth}
\centering
\includegraphics[width=0.9\linewidth]{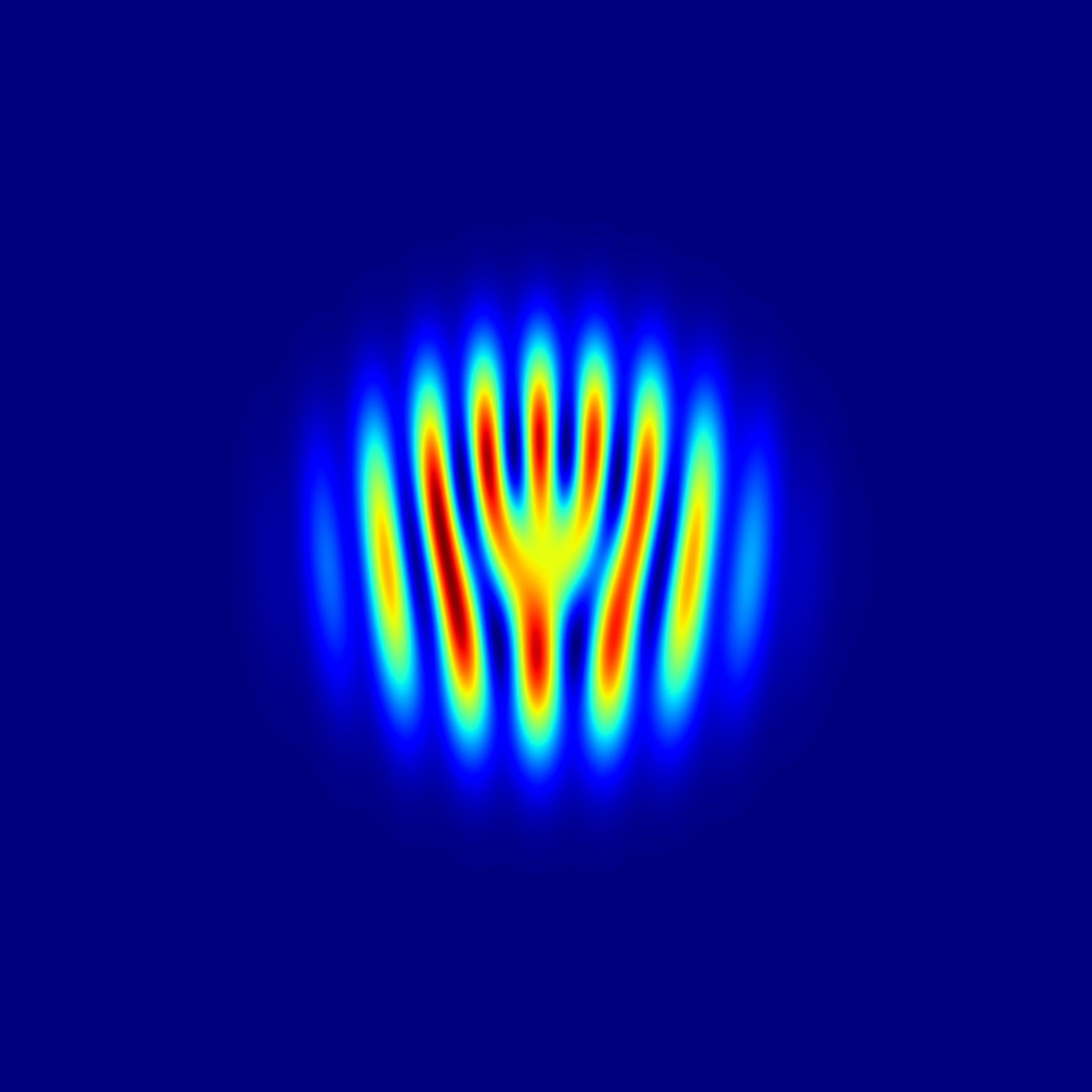} \\ $LG_{0,2}$
\end{minipage}
\hfill
\begin{minipage}[h]{0.24\linewidth}
\centering
\includegraphics[width=0.9\linewidth]{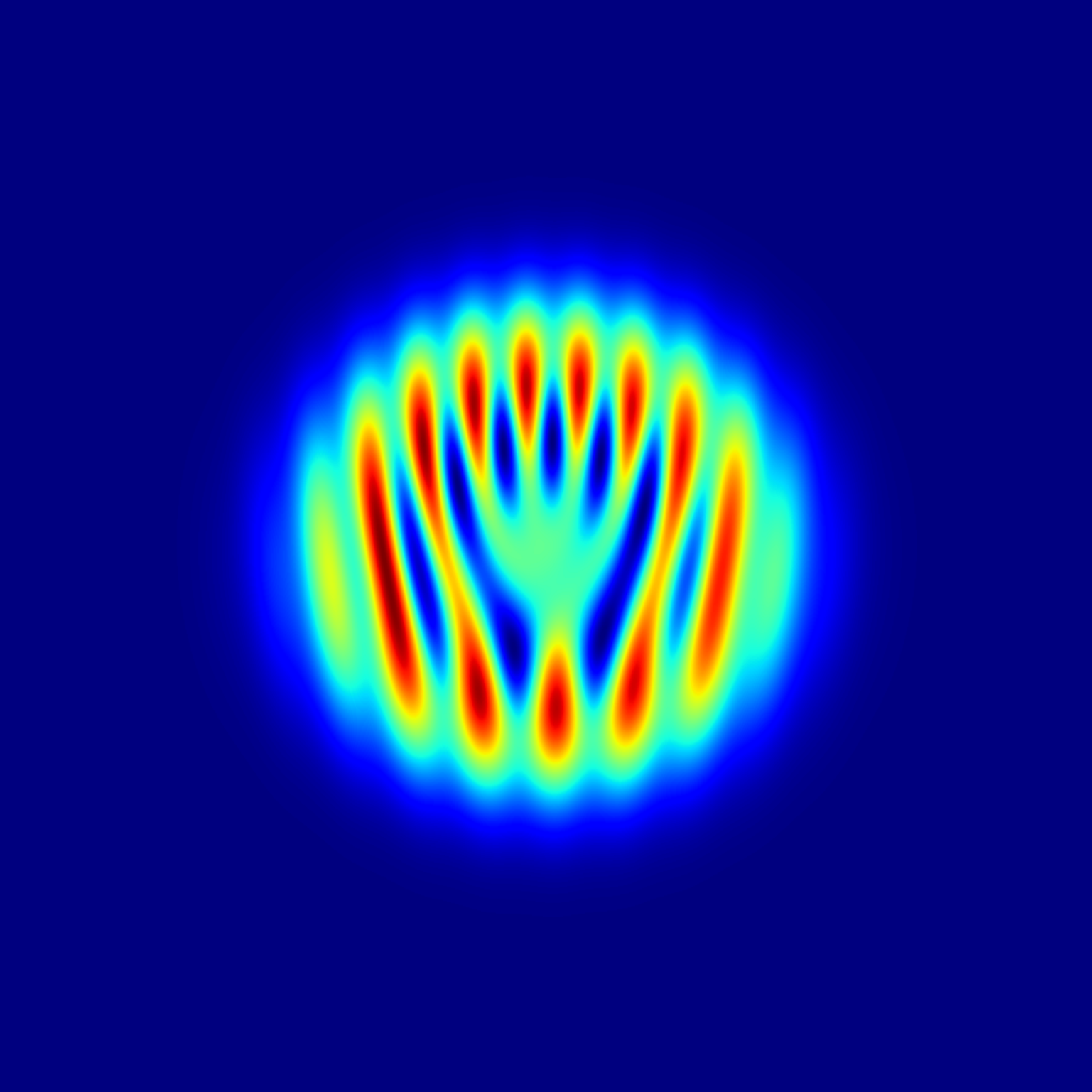} \\ $LG_{0,3}$
\end{minipage}
\hfill
\begin{minipage}[h]{0.24\linewidth}
\centering
\includegraphics[width=0.9\linewidth]{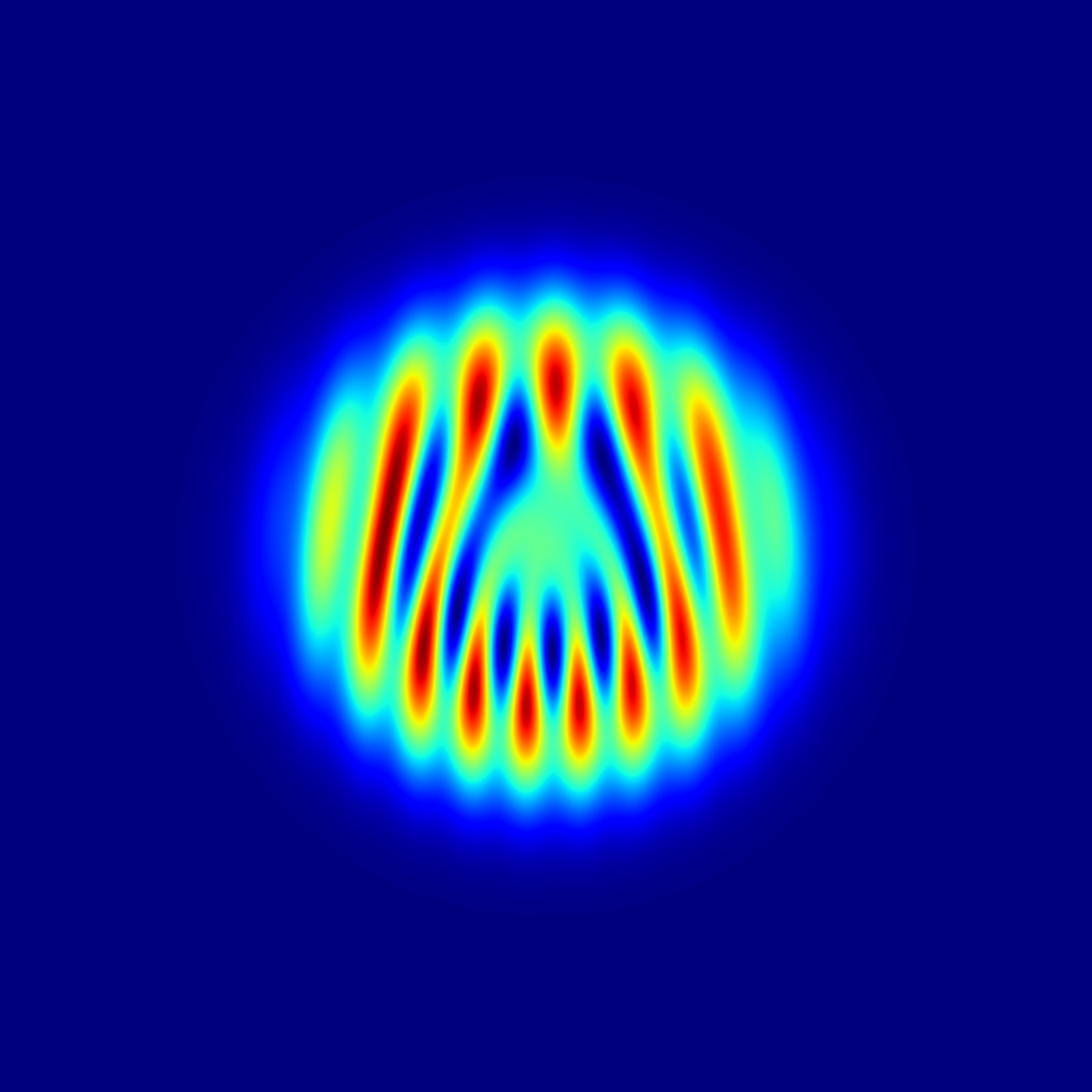} \\ $LG_{0,-3}$
\end{minipage}
\captionsetup{width=1.0\linewidth}
\caption{Numerical simulation of the interference pattern of LG modes with a Gaussian reference beam. The first index is the radial number $p$, the second is the azimuthal number $\ell$.}
\L{Fig1}
\end{figure}

However, it should be noted that this approach is not without its drawbacks. In the seminal works \cite{Basistiy-1995, Ricci-2012} was demonstrated that, in the context of LG beams with an OAM equal to $\ell$ propagating within a turbulent atmosphere, the central singularity undergoes a decay process into a set of $|\ell|$ singularities, situated on a circle of non-zero radius. These modes correspond to $\ell$ equal to either +1 or -1. Consequently, the interference pattern will demonstrate not one "fork" corresponding, for example, to the $LG_{0,3}$ mode, but three "forks" corresponding to the $LG_{0,1}$ mode. This negative aspect significantly complicates the task of multiplexing to increase the information capacity of the optical channel, since it becomes difficult, or even impossible, to distinguish the contribution from modes with different values of the OAM in one interference pattern.

Conversely, the utilisation of LG modes that vary in radial numbers evades this disadvantage. In the context of propagating in a turbulent medium, the ring structure persists and remains discernible in the interference pattern, a finding that is substantiated by simulation results.

The propagation of beams in turbulent atmospheres was modelled using the split-step beam propagation method \cite{Yousef-2020, Belmonte-2000}. This method enables the consideration of both beam diffraction and the effects of an inhomogeneous medium caused by turbulence (see Appendix).

The results of modelling the intensity of the LG beam and its interference with the Gaussian beam are shown in Fig. \ref{Fig2}.
\begin{figure}
\centering
\includegraphics[width=0.98\linewidth]{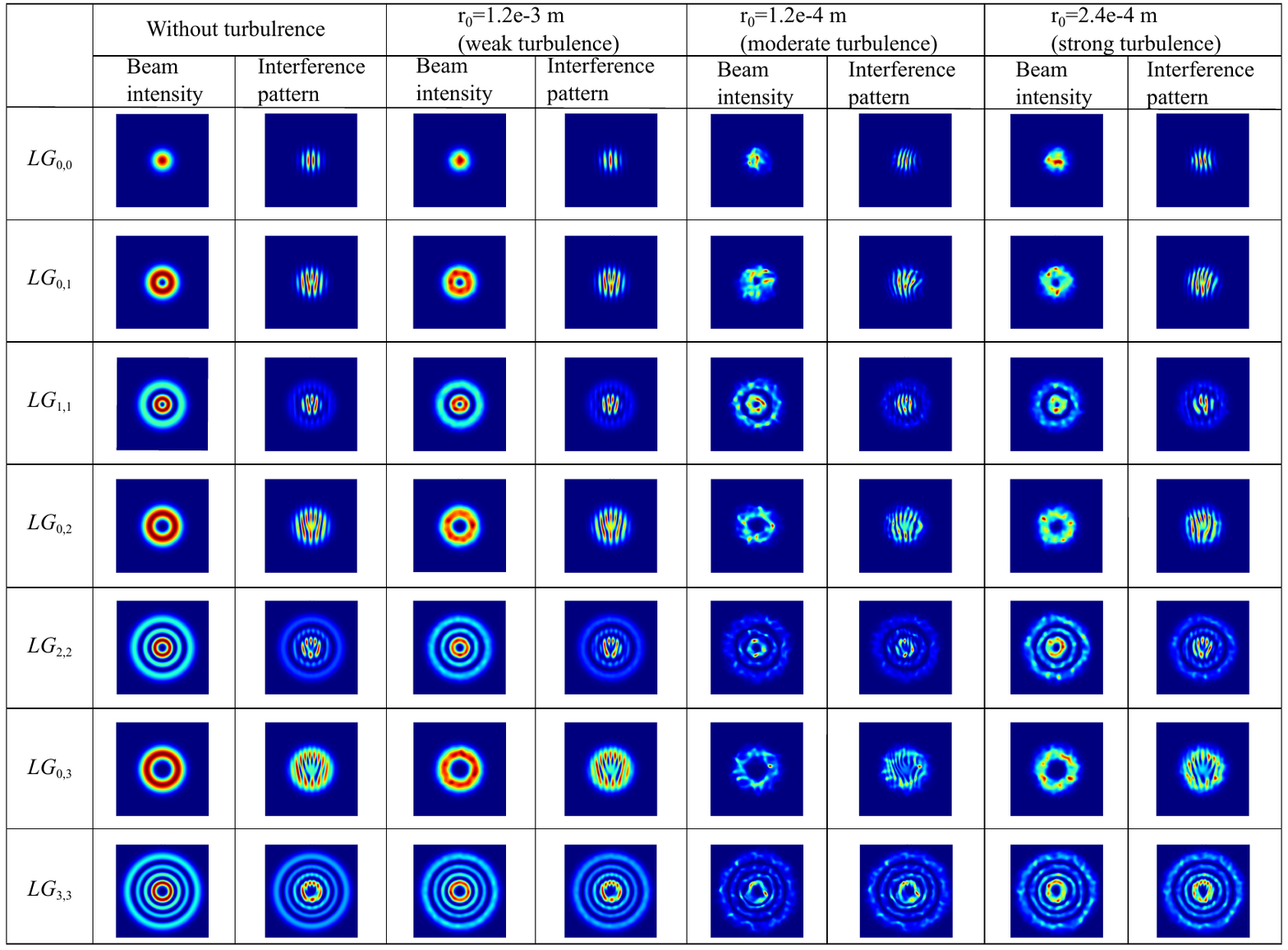}
\caption{Summary table of the results of numerical simulation of the passage of LG modes $LG_{0,0}, LG_{0,1}, LG_{1,1}, LG_{0,2}, LG_{2,2}, LG_{0,3}, LG_{3,3}$ through a turbulent gas medium; $r_0$ --- the Fried parameter characterizing the value turbulence of the medium (see Appendix). Images of the intensity of the transmitted beam and the result of interference with the reference Gaussian beam are presented.}
\L{Fig2}
\end{figure}

It is evident that in circumstances of moderate to strong turbulence, the identification of LG modes that vary in the azimuthal number, designated as $\ell$, becomes a challenging endeavour. This is due to the necessity of employing adaptive optics techniques, amongst other factors. Conclusive findings of a similar nature have been presented in the works \cite{Zhao-2016, Li-2017, Torner-2005}. In these works, the overlapping of modes with differing OAM propagating in a turbulent medium is analysed. It is evident that the transmission of information through an atmospheric optical channel using LG modes with different values of $\ell$, without the use of additional wavefront correction systems, becomes extremely sensitive to both channel quality and signal transmission distance.

Conversely, as the turbulence strength increases, the radial structure of the profile is distorted at a much slower rate in comparison to the profile structure determined by the value of the orbital angular momentum of a given beam. In this context, it appears more appropriate to employ the radial number $p$ in the context of discrete amplitude coding. This is due to the fact that, in this scenario, the detection process of the received mode is simplified and reduced to a straightforward calculation of the number of rings in the interference pattern or intensity profile of the received beam.

\section{Results: The relationship between Laguerre-Gaussian modes and Zernike polynomials}

Consider a complete set of Zernike polynomials orthogonal in a circle of radius $w_0$\cite{Wunsche-2005}:
\BE
Z_n^m(\rho,\varphi) = \sqrt{\frac{n+1}{\pi w_0^2}} R_n^m (\rho) \, e^{i m\varphi}, \L{2}
\EE
where
\BE
R_n^m (\rho) = \2! \sum_{t=0}^{(n-|m|)/2} \2! (-1)^t \frac{(n - t)!}{t!  \( \frac{n + |m|}{2} - t \)! \( \frac{n - |m|}{2} - t \)! } \( \frac{\rho}{w_0} \)^{n-2t}. \L{3}
\EE

Each Zernike polynomial provides a description of an aberration of a specific type and order. When an arbitrary wavefront is decomposed into a series by Zernike polynomials, the expansion coefficients demonstrate the contribution of one type or another of aberrations. The influence of aberrations on the propagation of spatially structured beams plays an extremely important role. For instance, the disintegration of the singularity at the centre of the LG beam may be attributed to its initial astigmatism \cite{Basistiy-1993}, which can result in a dynamic inversion of the OAM when the beam propagates in the medium \cite{Molina-2001}. In work \cite{Khorin-2018}, the behaviour of beams whose wavefront is a superposition of Zernike polynomials is investigated. It can thus be concluded that the decomposition of an arbitrary LG mode according to the basis of Zernike polynomials will allow for an estimation of the possible degree of distortion of the beam passing through a turbulent medium. This estimation is dependent upon the values of the radial $p$ and azimuthal $\ell$ numbers.

We will investigate the case of a weakly converging beam: at distances along the $z$ axis when the Rayleigh length is $z_R >> z$. This makes it possible to simplify the expression (\ref{1}) by reducing the ratio $z/z_R$ to zero. That is, we will consider the expansion of the transverse beam profile in the region of the constriction $z = 0$. Then in the case of substitution
\BE
x = \frac{2 \rho^2}{w_0^2} \L{4}
\EE
expression (\ref{1}) takes the form
\BE
LG_{p,\ell}(x, \varphi) =
\frac{\mathcal{N}_{p,\ell}}{w_0} \
x^{|\ell|/2} \, L^{|\ell|}_p [ x ] \ e^{-x/2} \, e^{i \ell \varphi}. \L{5}
\EE
The decomposition over the complete set of orthogonal Zernike polynomials can be represented as follows:
\BE
LG_{p,\ell}(x, \varphi) = \sum_{n=0}^{\infty} \sum_{m=-n}^{n} C_{nm}^{\, p,\ell} \,Z_n^m(x, \varphi), \L{6}
\EE
where the expansion coefficients can be found as
\BE
C_{nm}^{\, p,\ell} = \frac{w_0^2}{4} \! \int_{0}^{2\pi} \5! d\varphi \! \int_{0}^{2} \3! dx \, LG_{p,\ell}(x, \varphi) \, {Z_n^m}^{*}(x, \varphi). \L{7}
\EE
After angle variable integration and taking into account the Rodrigues formula for the Laguerre polynomial in binomial form \cite{Koornwinder-2010}
\BE
\BA
&C_{nm}^{\, p,\ell} =
\frac{\mathcal{N}_{p,\ell} \sqrt{\pi (n+1)}}{2} \sum_{k=0}^{p} \! \sum_{h=0}^{\frac{n-|m|}{2}} (-1)^{k+h}
\binom{p+|\ell|}{p-k} \binom{n-h}{\frac{n-|m|}{2}} \binom{\frac{n-|m|}{2}}{h}\frac{1}{k!}\cdot 2^{h - n/2}\cdot\\
&\cdot\int_0^2 \! dx \, x^{\frac{n + |\ell|}{2} + k - h} e^{-x/2} \, \delta_{m,l}. \L{8}
\EA
\EE
The integral over $dx$ is a well-known \cite{GradRyz}:
\BE
\int_0^{u} x^{\nu-1} e^{-\mu x} dx = \mu^{-\nu} \gamma \[ \nu,\mu\cdot u \], \quad Re(\nu) > 0, \L{9}
\EE
where $\gamma$ is an lower incomplete gamma function. Using the Kronecker symbol, as well as taking into account the explicit form of the normalization factor $\mathcal{N}_{p,\ell}$, we obtain the final expression for the expansion coefficients
\BE
\BA
&C_{n}^{\, p,\ell} =
\sqrt{ \frac{(n+1) \, 2^{|\ell|+1}  \, p!}{(p + |\ell|)!}} \,
\sum_{k=0}^{p} \sum_{h=0}^{\frac{n-|\ell|}{2}} \, (-1)^{k+h} \, \frac{2^k}{k!} \,
\binom{p+|\ell|}{p-k} \binom{n-h}{\frac{n-|\ell|}{2}} \binom{\frac{n-|\ell|}{2}}{h} \cdot\\
&\cdot\gamma \! \[\frac{n + |\ell|}{2} +1 + k - h, 1\], \L{10}
\EA
\EE
in this case, the expression (\ref{6}) will take the form
\BE
LG_{p,\ell}(x, \varphi) = \sum_{n=0}^{\infty} C_{n}^{\, p,\ell} \,Z_n^\ell(x, \varphi). \L{11}
\EE
\section{Discussion}

To assess how the magnitude and type of aberrations are affected by radial $p$ and azimuthal $\ell$ numbers, we compare the absolute values of the expansion coefficients $|C_{n}^{\,p,\ell}|$. In Fig. \ref{Fig3} two dependences of the coefficients $|C_{n}^{p,l}|$ are presented as functions of $\ell$ and $n$ for $p$ equal to 0 and 15. Both graphs show that under the condition
\BE
n < |\ell| \L{12}
\EE
the coefficients $|C_{n}^{p,\ell}|$ are zero. This result is valid for an arbitrary $p$. Thus, zero values of the coefficients determining astigmatism can reduce the possibility of a dynamic OAM inversion during propagation of the LG beam \cite{Basistiy-1993, Molina-2001}.
\begin{figure}[h!]
\begin{minipage}[h]{0.45\linewidth}
\centering
\includegraphics[width=1\linewidth]{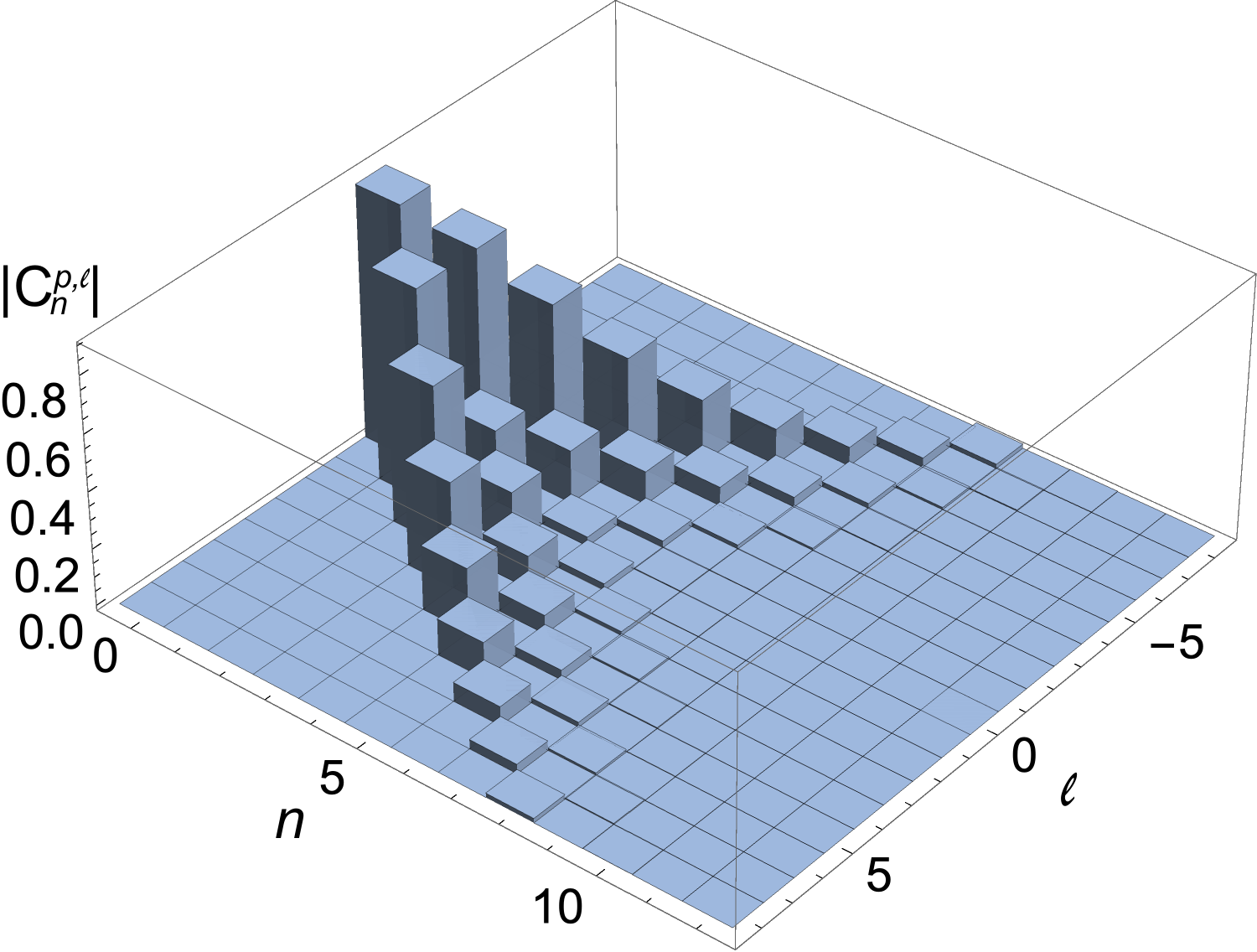} \\ (a) $p=0$
\end{minipage}
\hfill
\begin{minipage}[h]{0.45\linewidth}
\centering
\includegraphics[width=1\linewidth]{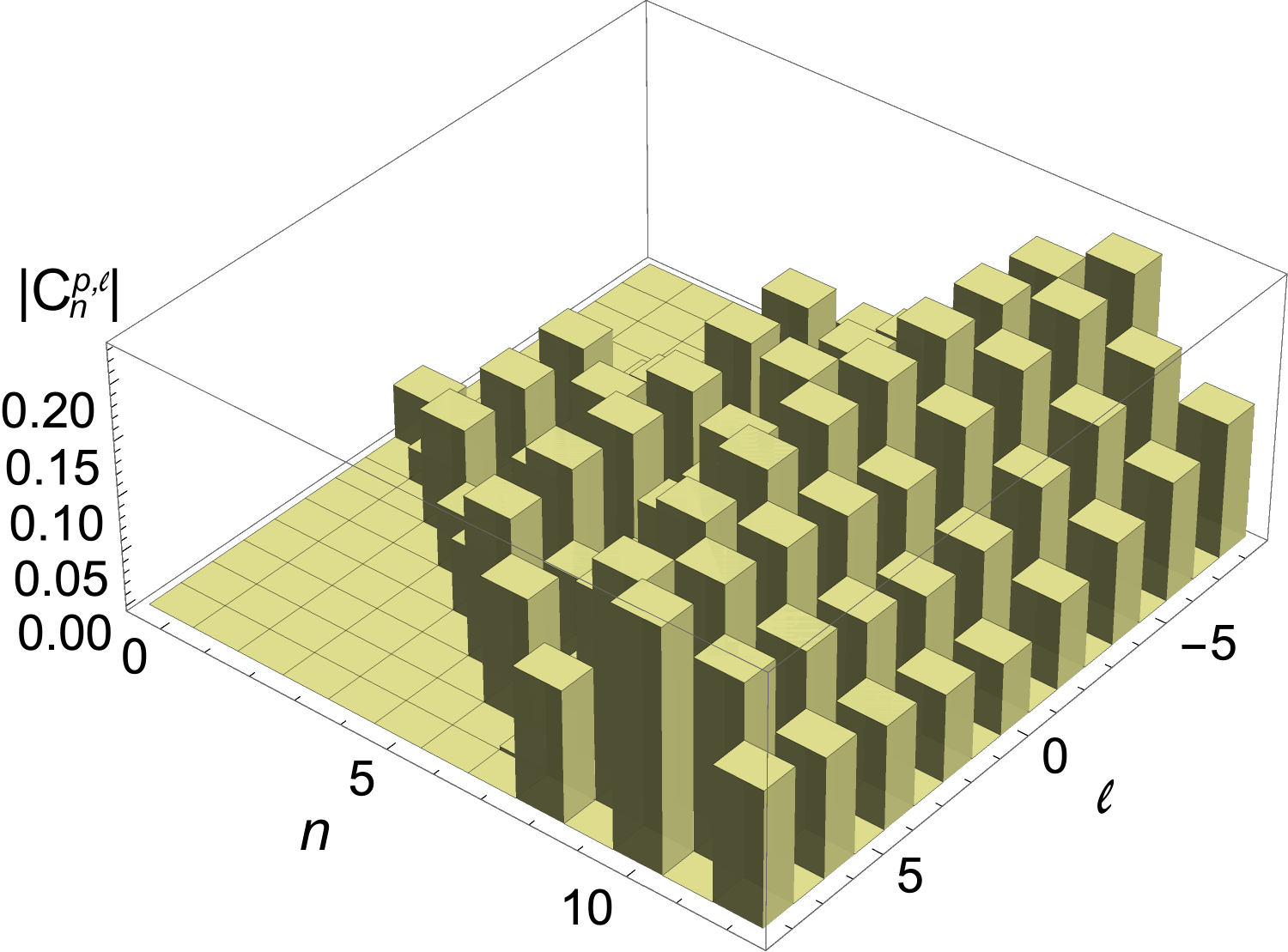} \\ (b) $p=15$
\end{minipage}
\caption{The absolute values of the coefficients of expansion in a series of Zernike polynomials $|C_{n}^{p,\ell}|$ for fixed values of radial numbers $p$.}
\L{Fig3}
\end{figure}

In Fig. \ref{Fig4} the dependences of the absolute values of the coefficients $|C_{n}^{p,l}|$ are presented as functions of the radial number $p$ and the Zernike polynomials index $n$ for fixed azimuthal numbers $\ell$. It is evident that for any given value of $n$ and $\ell$ with an increase in $p$, the evolution of the coefficients is oscillatory, while the oscillation amplitudes exhibit a general tendency to decrease. As illustrated in Fig. \ref{Fig4} (b): It is evident that the coefficients with index values $n < 5$, have a value of zero. This result is consistent with the inequality given in equation (\ref{12}). It is also noteworthy that this phenomenon is illustrated in Figure 4. It can be demonstrated that all coefficients with an even value of n are zero, and vice versa, in Fig. \ref{Fig4} (a). Furthermore, it is evident that all coefficients with an odd value of n are zero. This phenomenon can be elucidated by referring to the rule for the values of the index n \cite{Born-1964}. According to this rule, the difference $n - |m|$ (where $m$ denotes $\ell$ in this context) must be an even number. This principle is exhibited in the upper index of summation in expressions (\ref{3}), (\ref{8}), (\ref{10}).
\begin{figure}[h!]
\begin{minipage}[h]{0.45\linewidth}
\centering
\includegraphics[width=1\linewidth]{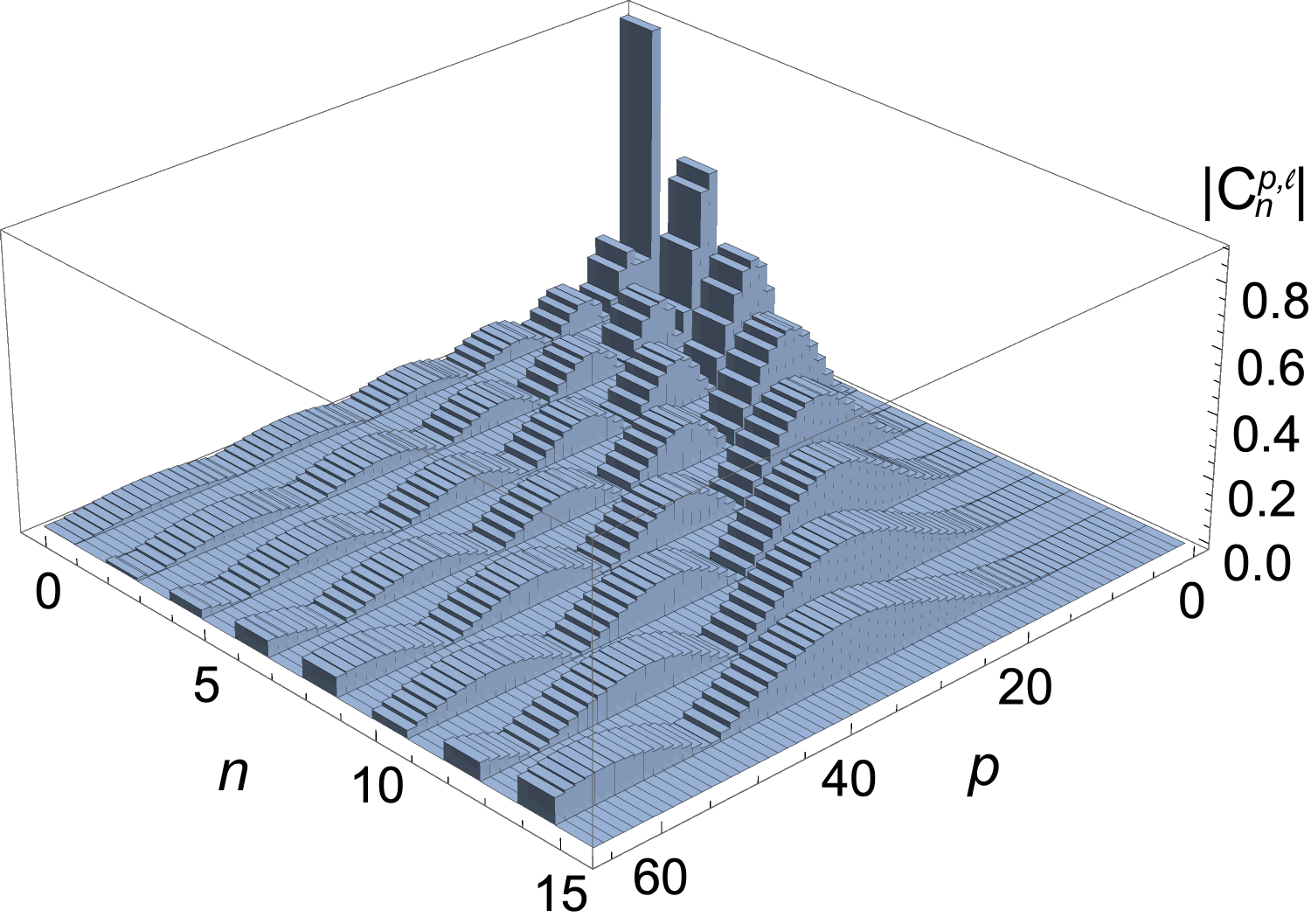} \\ (a) $\ell=0$
\end{minipage}
\hfill
\begin{minipage}[h]{0.45\linewidth}
\centering
\includegraphics[width=1\linewidth]{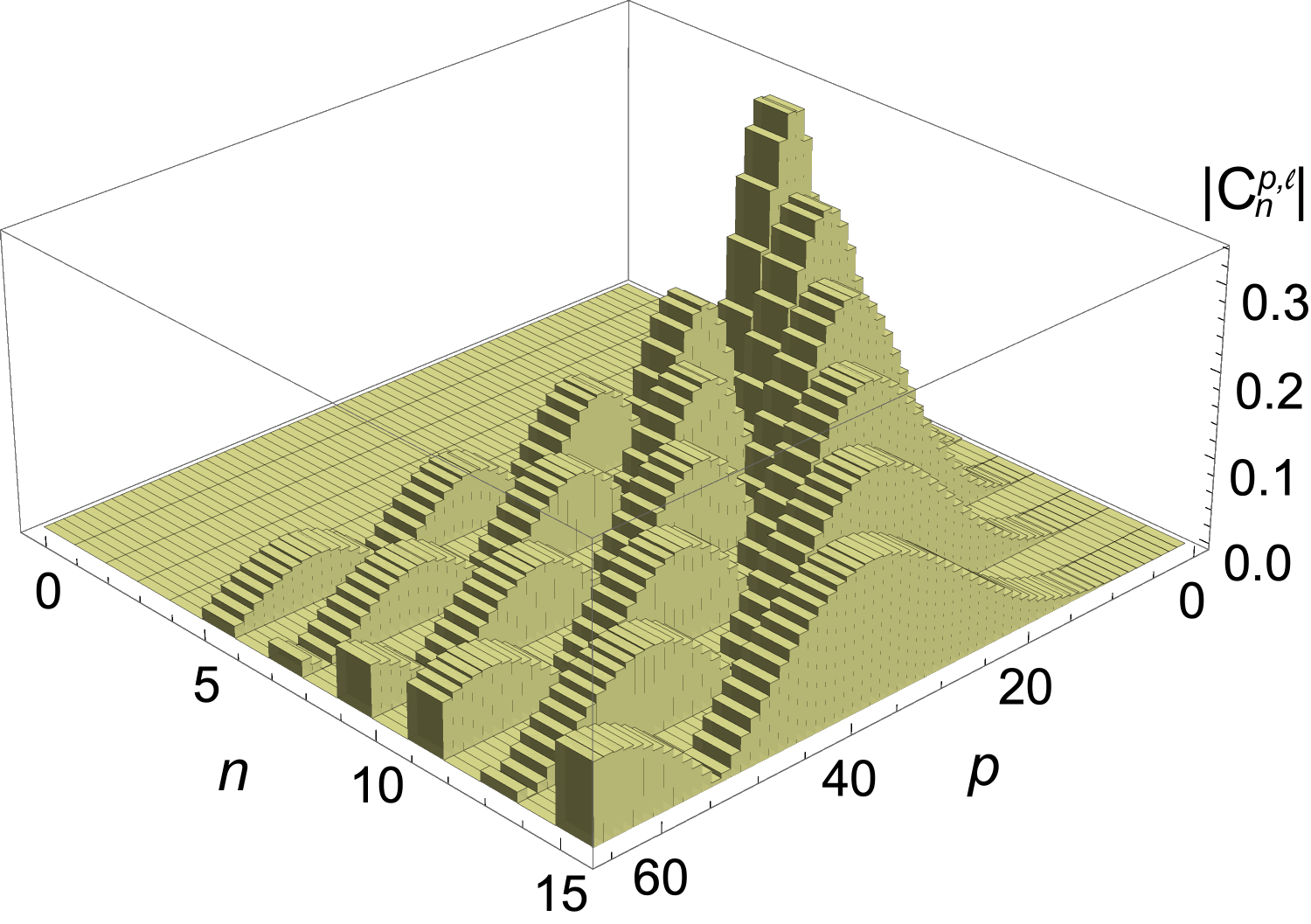} \\ (b) $\ell=\pm 5$
\end{minipage}
\caption{The absolute values of the coefficients of expansion in a series of Zernike polynomials $C_{n}^{p,\ell}$ for fixed values of the azimuthal number $\ell$.}
\L{Fig4}
\end{figure}

In \cite{Khorin-2018}, it was shown that modes whose wavefront expansion contains non-zero coefficients with matching indices $|\ell| = n$ (that is, $C_0^{p,0}$, $C_1^{p,1}$, etc.), are least resistant to the passage of turbulent media. However, in Fig. \ref{Fig5} it can be seen that there are such values of $p$ at which these coefficients are close or even exactly zero. Thus, by reducing the contribution of unstable Zernike modes, it is possible to achieve greater stability of the LG beams.
\begin{figure}[h!]
\centering
\includegraphics[width=0.5\linewidth]{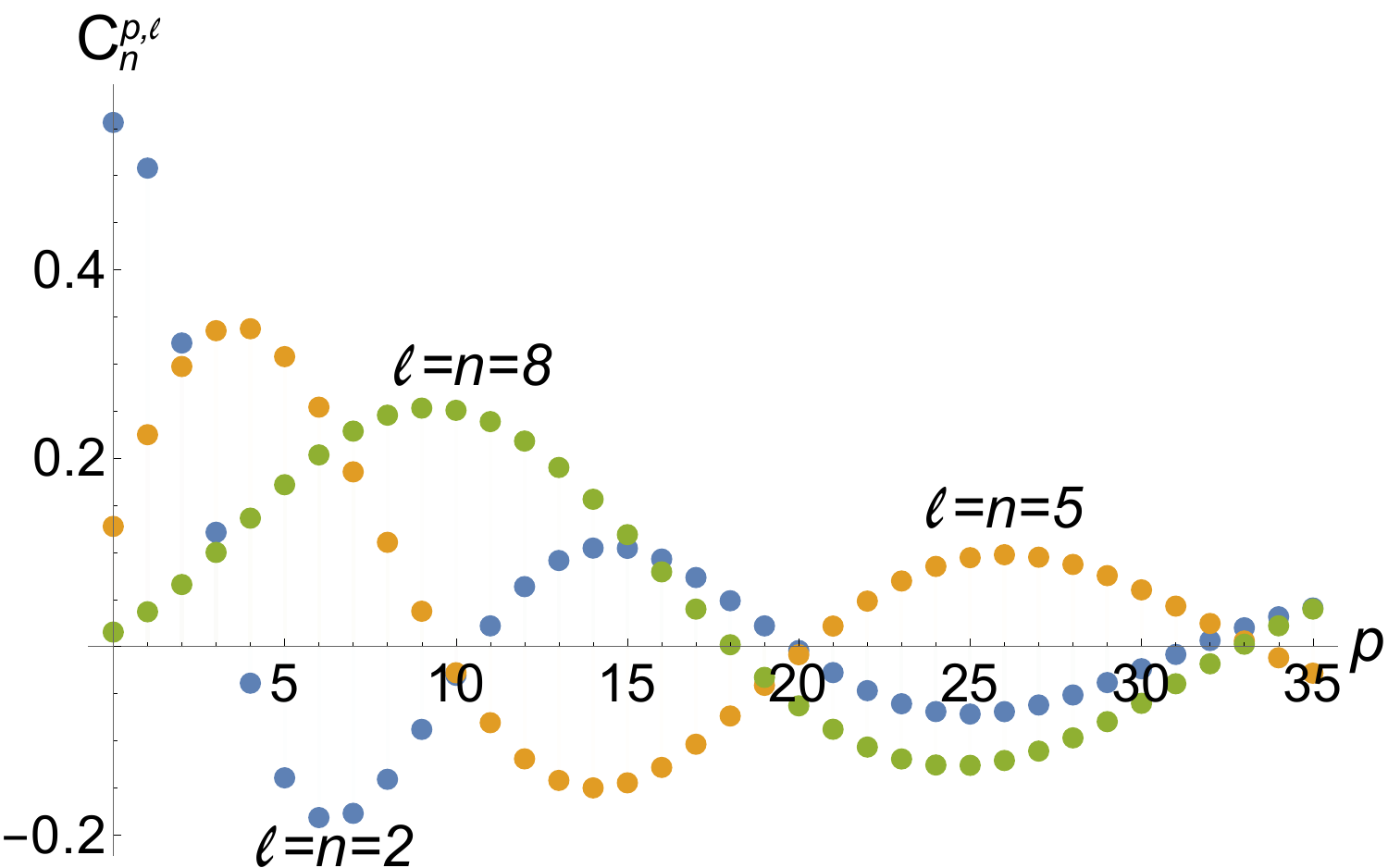}
\caption{Coefficients of the Zernike polynomial expansion $C_{n}^{p,\ell}$ as functions of $p$ in cases where the azimuthal number $\ell$ is equal to the index $n$.}
\L{Fig5}
\end{figure}
\section{Conclusion}

It has been demonstrated that there are discrepancies in the resistance to propagation in a turbulent gas medium of LG modes with different radial $p$ and azimuthal $\ell$ numbers. The simulation demonstrates the efficacy of employing a radial number for the encoding of information during the detection of a transverse profile utilizing a CCD camera.

It can be concluded from the expression for the coefficients of decomposition of LG functions obtained according to the basis of Zernike polynomials, and the graphical representation of these coefficients as a function of the numbers $p$ and $\ell$ , that the selection of an appropriate radial number has the capacity to improve the stability of the transverse profile of LG modes in terms of propagation through a turbulent medium. A general tendency is exhibited to decrease the absolute values of the coefficients $|C_n^{p,\ell}|$ with an increase in $p$, including coefficients for polynomials with matching indices. That is, to increase noise immunity, it would be advisable to use LG modes with large radial numbers. However, an increase in $p$ leads to an increase in the diameter of the beam, which is then limited by the aperture of the detecting device. However, when encoding on a selected range of azimuth numbers, it is possible to select the value (or values) of the radial number so that the overall negative effect of the passage of the LG beam through the atmosphere will be reduced.

\subsection*{Acknowledgments}
The authors declare that they have no conflict of interest.
\\

\appendix    

\section{Mathematical description of the method used for numerical modeling}
\label{sect:misc}
To increase the accuracy, the split step beam propagation method of the second order was used. Mathematically, this method can be described as follows. Consider the evolution of a beam with a profile defined by the function $E(x,y,z)$ between points in the space $z$ and $z+\Delta z$:
\begin{itemize}
    \item Let's move to k-space and take into account the diffraction of the beam when propagating between the points $z+\frac{\Delta z}{2}$:
    \BE
    \widetilde{E}\(k_x, k_y, z+\frac{\Delta z}{2}\)=\mathcal{F}\[E(x,y,z)\]\cdot e^{\frac{i\(k_x^2+k_y^2\)\(\frac{\Delta z}{2}\)}{2 k_0}},
    \EE
    where $\mathcal{F}\[.\]$ --- is a Fourier transform operation, and  $k_0=\frac{2\pi}{\lambda}$ --- is a wave number.

    \item Let's go back to physical space and take into account random processes of interaction with a turbulent medium by multiplying the beam profile by the operator $e^{-i \hat{S}\Delta z}, \hat{S}=\Delta n\cdot k_0$ and $\Delta n$ is a two-dimensional distribution of refractive index fluctuations:
    \BE
    E'\(x,y,z+\frac{\Delta z}{2}\)=\mathcal{F}^{-1}\[\widetilde{E}\(k_x,k_y,z+\frac{\Delta z}{2}\)\]\cdot e^{-i\hat{S}\Delta z}.
    \EE

    \item Let's take into account the rest of the diffraction effects:
    \BE
    \widetilde{E}\(x,y,z+\Delta z\)=\mathcal{F}\[E'\(x,y,z+\frac{\Delta z}{2}\)\]\cdot e^{\frac{i\(k_x^2+k_y^2\)\(\frac{\Delta z}{2}\)}{2 k_0}}.
    \EE

    \item The final expression for the field profile at the point $z+\Delta z$ is obtained by means of the inverse Fourier transform:
    \BE
    \BA
    &E(x,y,z+\Delta z)=\mathcal{F}^{-1}\[\widetilde{E}\(x,y,z+\Delta z\)\]=\\
    &=\mathcal{F}^{-1}\[\mathcal{F}\[\mathcal{F}^{-1}\[\mathcal{F}\[E(x,y,z)\]\cdot e^{\frac{i\(k_x^2+k_y^2\)\(\frac{\Delta z}{2}\)}{2 k_0}}\]\cdot e^{-i\hat{S}\Delta z}\]\cdot e^{\frac{i\(k_x^2+k_y^2\)\(\frac{\Delta z}{2}\)}{2 k_0}}\].
    \EA
    \EE
\end{itemize}

\bibliographystyle{iopart-num}
\bibliography{report.bib}   


\end{document}